\documentclass[runningheads]{llncs}
\usepackage[T1]{fontenc}
\usepackage{graphicx}
\usepackage{amsmath}
\usepackage{booktabs} 
\usepackage{subcaption}
\usepackage{xcolor}

\usepackage{hyperref}
\usepackage{color}

\begin{document}
\title{CROCODILE \includegraphics[height=1em]{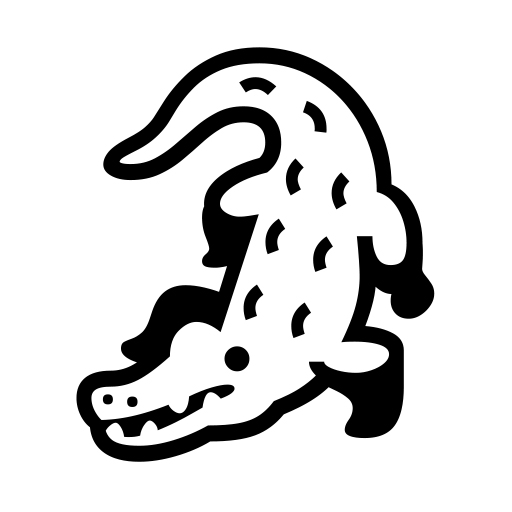}: Causality aids RObustness via COntrastive DIsentangled LEarning}

\titlerunning{Causality aids RObustness via COntrastive DIsentangled LEarning}
%
\author{
Gianluca Carloni\inst{1,2}\orcidID{0000-0002-5774-361X} \and
Sotirios A. Tsaftaris\inst{3}\orcidID{0000-0002-8795-9294} \and
Sara Colantonio\inst{1}\orcidID{0000-0003-2022-0804}
}
\authorrunning{G. Carloni et al.}
\institute{
National Research Council, Pisa, Italy \\
\email{gianluca.carloni@isti.cnr.it} \and
University of Pisa, Pisa, Italy \and
The University of Edinburgh, Edinburgh, UK
}
\maketitle              
\begin{abstract} 
Due to domain shift, deep learning image classifiers perform poorly when applied to a domain different from the training one. For instance, a classifier trained on chest X-ray (CXR) images from one hospital may not generalize to images from another hospital due to variations in scanner settings or patient characteristics.
In this paper, we introduce our CROCODILE framework, showing how tools from causality can foster a model's robustness to domain shift via feature disentanglement, contrastive learning losses, and the injection of prior knowledge. This way, the model relies less on spurious correlations, learns the mechanism bringing from images to prediction better, and outperforms baselines on out-of-distribution (OOD) data.
We apply our method to multi-label lung disease classification from CXRs, utilizing over 750000 images from four datasets.
Our bias-mitigation method improves domain generalization and fairness, broadening the applicability and reliability of deep learning models for a safer medical image analysis.
Find our code at: \href{https://github.com/gianlucarloni/crocodile}{https://github.com/gianlucarloni/crocodile}. 

\end{abstract}
\section{Introduction}
Domain shift bias is the problem of machine learning (ML) models performing not consistently across \textit{in-distribution} (ID) and \textit{out-of-distribution} (OOD) data. The former are independent and identically distributed (i.i.d) to the data on which the model was trained. Conversely, data are OOD when their distribution essentially differs from the source one, such as chest X-rays (CXR) coming from a different hospital than the training one \cite{pooch2020can,cohen2020limits,zhang2022learning}.
Traditional ML models still tend to rely on spurious correlations seen during training for predicting the outcome and spectacularly fail when those shortcut associations are not present in OOD data, for instance, due to variations in scanner settings, image artifacts, or patient demographics \cite{castro2020causality,sanchez2022causal,bercean2023breaking,hartley2023neural}. For this reason, the field of domain generalization (DG) has searched for ways to make deep learning (DL) models learn robust features that could generalize better to unseen domains \cite{li2021domain,ouyang2022causality,wang2022generalizing,zunaed2024learning}.

Conceptually, we could think of a set of features that causally determine the outcome and are invariant to shifts in non-relevant attributes, as well as a separate set of features that are spuriously correlated with the outcome but do not have a causal effect. Some works have proposed using tools from causal inference to achieve this disentanglement \cite{wang2021causal,sui2022causal,nie2023chest}. The common idea is that using the causal instead of the spurious features would allow a model to learn the underlying mechanism and be more robust on new data.
However, these efforts try to model domain shifts implicitly, with a scope limited to the disease prediction task, disregarding the wealth of information on possible domain shifts from different source data sets.

In this work, we advance this causal/spurious feature disentanglement on a cross-domain level by leveraging information from different datasets in a contrastive learning setting. We conceive a domain prediction branch along the disease-prediction branch to instill domain awareness into the model's representations. Moreover, we propose a new way to inject background medical knowledge, effectively designing a task prior to guiding learning and fostering DG.

\section{Methodology}
\begin{figure}[t]
\includegraphics[width=\textwidth]{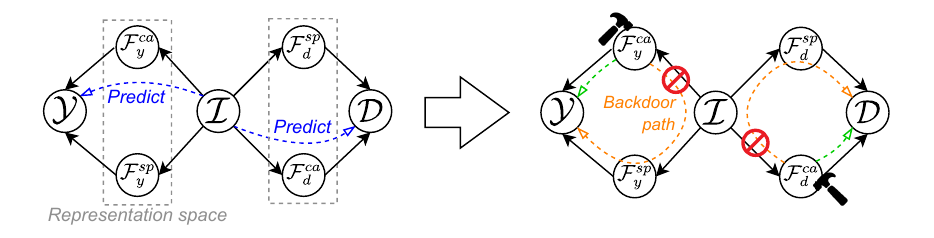}
\caption{A causal view on classifying medical images $\mathcal{I}$ coming from different domains $\mathcal{D}$ for the presence of diseases $\mathcal{Y}$. By applying the latent causal intervention (hammer), the backdoor path through the spurious features is cut off.}
\label{fig:introduction}
\end{figure}
We define a structural causal model (SCM) \cite{pearl2010causal} for medical image classification in Fig \ref{fig:introduction}.
Given the input images $\mathcal{I}$, such as CXRs, and the disease classification $\mathcal Y$, we obtain two sets of features via feature extraction.
We denote $\mathcal F^{ca}_y$ the causal features that truly determine the outcome (e.g., the patchy airspace opacification typical in pneumonia).
Similarly, we denote $\mathcal F^{sp}_y$ the spurious features, determined by data bias's confounding effect, which are unrelated to a disease (e.g., metal tokens on the image corners).
Ideally, $\mathcal Y$ should be caused only by $\mathcal F^{ca}_y$, but is naturally confounded by $\mathcal F^{sp}_y$, as both types of features 
usually coexist in medical data. Unfortunately, conventional models tend to learn the correlation $P(\mathcal{Y}|\mathcal{F}^{ca}_y)$ via the shortcut (backdoor) path $\mathcal{F}^{ca}_y \leftarrow \mathcal{I} \rightarrow \mathcal F^{sp}_y \rightarrow \mathcal Y$ instead of the desired $\mathcal F^{ca}_y \rightarrow \mathcal Y$. As we detail next, we exploit the \textit{do-calculus} from causal theory \cite{pearl2014interpretation} on the causal features to block the backdoor path, estimating $P(\mathcal{Y}|do(\mathcal{F}^{ca}_y))$.
Following the same idea, we conceive two other sets of features extracted from $\mathcal{I}$, this time concerning the trivial task of predicting from which source domain come the data $\mathcal{D}$: $\mathcal F^{ca}_d$ would be the features that are relevant to distinguish different domains, and $\mathcal F^{sp}_d$ the confounding features. 

\subsection{Disease-branch and Domain-branch}
\begin{figure}
\includegraphics[width=\textwidth]{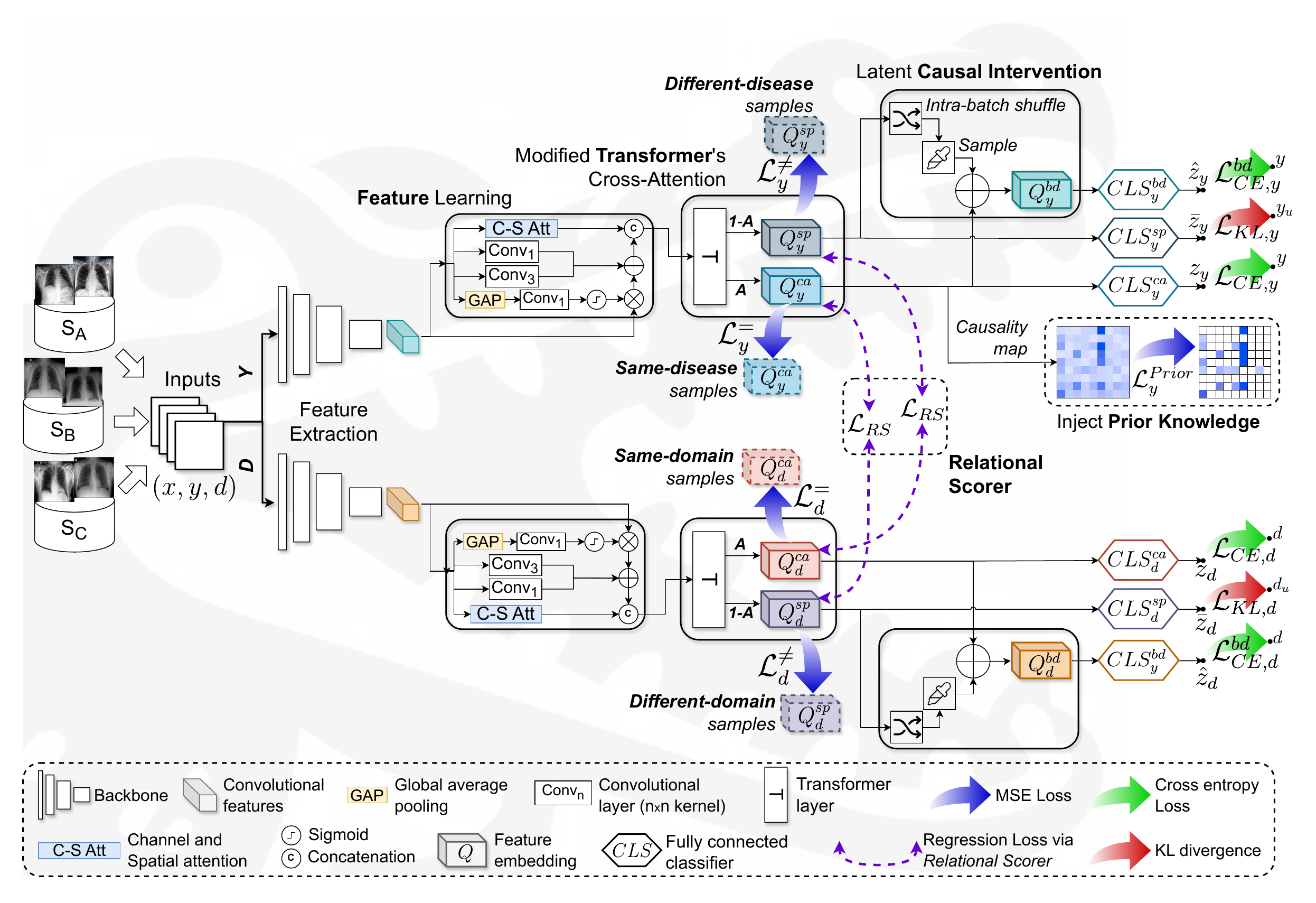}
\caption{CROCODILE involves two branches to learn robust, invariant features for predicting the labels from medical images (e.g., multi-label findings from CXRs) while disregarding confounding features.
We disentangle \textit{causal} features determining the label from \textit{spurious} features associated with the label due to domain shift.
We exploit images from multiple domains in a contrastive learning scheme and propose a new way to inject prior knowledge. Best seen in color.}
\label{fig:crocodile_framework}
\end{figure}

We present our overall framework in Fig \ref{fig:crocodile_framework}. 
A \textit{disease prediction} branch learns to extract useful image features to predict the medical finding (e.g., pneumothorax or atelectasis in a CXR), regardless of the different domains. On another parallel branch for \textit{domain prediction}, the image features that are useful for the trivial task of predicting the domain the images come from are learned (regardless of the different diseases).
The architecture is trained end-to-end.
Each branch involves a feature extraction backbone
followed by
a block to enhance features
via channel- and spatial- attention \cite{pan2022integration}.
Then, a Transformer network \cite{vaswani2017attention} with a modified cross-attention mechanism yields not only the usual set $\mathbf{A}$ of attention scores relevant to the task but also its complementary set $\mathbf{1}-\mathbf{A}$, thus encoding disentangled causal and spurious feature embeddings, $Q^{ca}$ and $Q^{sp}$, respectively.
Finally, three classifiers connect the features $Q$ to the classification logits $z$.
In the following sections, we design specific contrastive learning losses and introduce a novel way to inject prior knowledge about the medical task.

\subsection{Feature Disentanglement and Causal Intervention}
For each branch, we need to make $Q^{ca}$ and $Q^{sp}$ capture the authentic and trivial aspects from the input samples. 
To achieve the correctness of the predictions, we impose two cross-entropy (CE) loss terms, $\mathcal L_{CE,y}$ and $\mathcal L_{CE,d}$, over the classification logits $z_y$ and $z_d$ from the causal features $Q^{ca}_y$ and $Q^{ca}_d$, supervised by the disease labels $y$ and domain labels $d$, respectively.
To make $Q^{sp}$ features encode the trivial patterns that are unnecessary for classification, we push its predictions $\bar{z}_y$ and $\bar{z}_d$ evenly to all respective categories. We define the uniform classification losses $\mathcal L_{KL,y}$ and $\mathcal L_{KL,d}$ as the KL-divergence between the spurious features and the respective uniform distribution ($y_u$ or $d_u$).
To alleviate the confounding effect, we implement the backdoor adjustment by performing a latent causal intervention \cite{sui2022causal,nie2023chest}: we stratify the spurious features appearing from training data and pair the causal set of features with those stratified spurious features to compose the \textit{intervened} graph. This way, we fit the concept of \textit{borrowing from others} (i.e., "if everyone has it, it is as if no one has it").
We impose CE losses $\mathcal L_{CE,y}^{bd}$ and $\mathcal L_{CE,d}^{bd}$ between the logits $\hat{z}_y$ and $\hat{z}_d$ obtained from the corresponding intervened features $Q^{bd}$ and the same ground-truth label for the causal features. This way, we push the predictions of such intervened images to be invariant and stable across different stratifications due to shared causal features. Practically, we approximate this operation with an intra-batch shuffling of $Q^{sp}$ followed by random sampling (with $0.3$ drop probability) and addition to $Q^{ca}$.
By combining the supervised CE loss, the KL loss, and the backdoor CE loss for each branch, we obtain the two following equations:

\begin{equation}
\label{eq:L_y}
\mathcal{L}_y = - (
    \lambda_1
    \underbrace{y^\top\!\!\log (z_y)}_{\mathcal{L}_{CE,y}}
    +
    \lambda_2
    \underbrace{KL(y_{u}, \bar{z}_y)}_{\mathcal{L}_{KL,y}}
    +
    \lambda_3
    \underbrace{y^\top\!\!\log (\hat{z}_y)}_{\mathcal{L}_{CE,y}^{bd}}
    )
\end{equation}
\begin{equation}
\label{eq:L_d}
\mathcal{L}_d = - (
    \lambda_4
    \underbrace{d^\top \log (z_d)}_{\mathcal{L}_{CE,d}}
    +
    \lambda_5
    \underbrace{KL(d_{u}, \bar{z}_d)}_{\mathcal{L}_{KL,d}}
    +
    \lambda_6
    \underbrace{d^\top \log (\hat{z}_d)}_{\mathcal{L}_{CE,d}^{bd}}
    )
\end{equation}

\subsection{Contrastive Learning}
\begin{figure}[t]
\includegraphics[width=\textwidth]{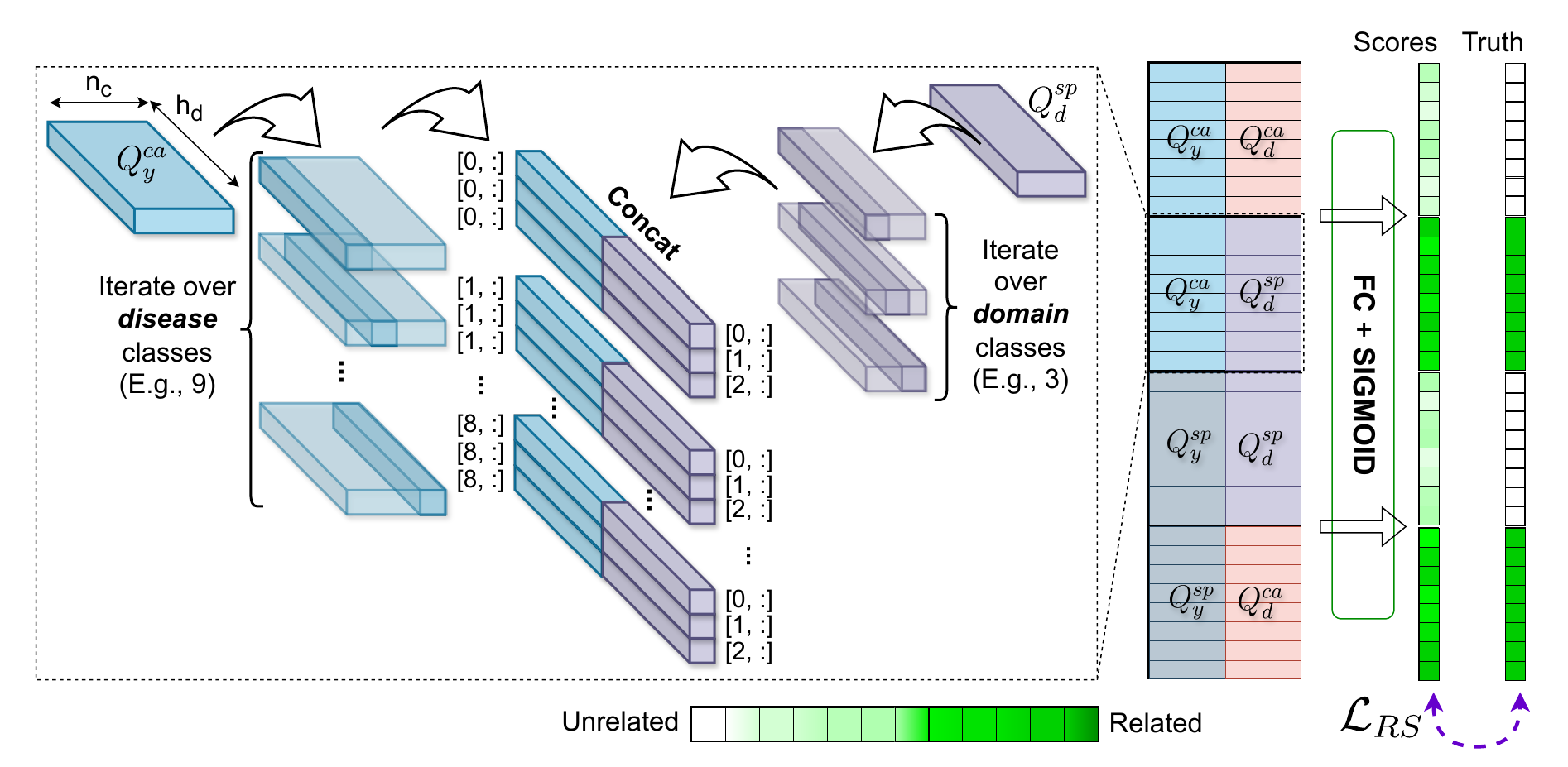}
\caption{Our \textit{Relational Scorer} stratifies and concatenates every combination of \textit{causal} and \textit{spurious} features across both tasks. With a fully connected layer and a consecutive sigmoid($\cdot$), it maps each pair to a \textit{relational score} between 0 and 1. We use an MSE loss regressing the relational scores to the ground truth. The model \textit{learns to compare} the four sets of disentangled features. Best in color.
}
\label{fig:relational_scorer}
\end{figure}

To attain cross-domain robustness, we posit there should also exist an alignment between the \textit{causal} features that determine the \textit{disease} and the \textit{spurious} features for the \textit{domain} prediction task. And the converse should also be true.
For instance, we want the regions of the image that determine the presence of pneumonia to be unrelated to what contributes to discerning different domains (e.g., spurious metal tokens). Conversely, the image aspects determining which domain the image comes from should be unrelated to what determines disease prediction.
However, we are interested in measuring the \textit{relational} alignment rather than the structural similarity of the representations. Matched (mismatched) pairs should "inform" ("repel") each other. Therefore, inspired by the concept of \textit{learning to compare} \cite{sung2018learning,cao2022learning}, we design a new module named \textbf{Relational Scorer} (RS) to learn which image representations' pairings are semantically related and which are not (Fig \ref{fig:relational_scorer}).
Our RS stratifies and combines each possible cross-branch pairing $p\in P=\{Q_y^{ca} \times Q_d^{ca} \cup Q_y^{ca} \times Q_d^{sp}\cup Q_y^{sp} \times Q_d^{sp} \cup Q_y^{sp} \times Q_d^{ca} \}$ and then maps them to a \textit{relational score} between $0$ and $1$.
We use an MSE loss regressing the relational scores $r$ to the ground truths $r^{GT}$: matched pairs have a similarity of $1$, and the mismatched pair have a similarity of $0$.
Although this problem may seem to be a \textit{classification} problem with label space \{$0$, $1$\}, we are predicting relation scores, which can be considered a \textit{regression} problem (with $r^{GT} \in \{0, 1\}$ generated by construction).
We set the ground truth to $1$ for the $Q_y^{ca}$-$Q_d^{sp}$ and $Q_y^{sp}$-$Q_d^{ca}$ pairings, and $0$ otherwise.
The resulting regression loss term is:
\begin{equation}
    \mathcal{L}_{RS}= - \lambda_7 \sum_{i=1}^{|P|}(r_i-r_i^{GT})^2
    \label{eq:L_RS}
\end{equation}

Moreover, we conceive other loss terms to enforce consistency/separation of medical image representations in a contrastive setting at a \textit{batch} level:
\begin{itemize}
    \item $\mathbf{\mathcal L_y^=}$: samples exhibiting a \textbf{common} radiological \textbf{finding} should lie close in \textit{disease-causal} feature space $Q_y^{ca}$, regardless of the source domain.
    \item $\mathbf{\mathcal L_y^{\neq}}$: samples exhibiting \textbf{different} radiological \textbf{findings} should lie close in \textit{disease-spurious} feature space $Q_y^{sp}$, regardless of the source domain.
    \item $\mathbf{\mathcal L_d^=}$: samples from the \textbf{same dataset} should lie close in \textit{domain-causal} feature space $Q_d^{ca}$, regardless of the diseases.
    \item $\mathbf{\mathcal L_d^{\neq}}$: samples from \textbf{different datasets} should lie close in \textit{domain-spurious} feature space $Q_d^{sp}$, regardless of the diseases.
\end{itemize}

We implement each of such terms via an MSE loss between the representation $Q$ of each sample in the batch and the corresponding average representation $\Tilde{Q}$ of samples with the same/different label:
\begin{equation}
\label{eq:L_y_batch}
\mathcal{L}_y^{batch} = - (
    \lambda_8
    \underbrace{\sum_{y \in \mathcal{Y}}(Q_y^{ca}-\Tilde{Q}_y^{ca})^2
    }_{\mathcal{L}_y^=}
    +
    \lambda_9
    \underbrace{\sum_{y \in \mathcal{Y}}(Q_y^{sp}-\Tilde{Q}_{not(y)}^{sp})^2
    }_{\mathcal{L}_y^{\neq}}
    )
\end{equation}
\begin{equation}
\label{eq:L_d_batch}
\mathcal{L}_d^{batch} = - (
    \lambda_{10}
    \underbrace{\sum_{d \in \mathcal{D}}(Q_d^{ca}-\Tilde{Q}_d^{ca})^2
    }_{\mathcal{L}_d^=}
    +
    \lambda_{11}
    \underbrace{\sum_{d \in \mathcal{D}}(Q_d^{sp}-\Tilde{Q}_{not(d)}^{sp})^2
    }_{\mathcal{L}_d^{\neq}}
    )
\end{equation}
\noindent where $\mathcal{Y}$ and $\mathcal{D}$ are the possible disease and domain labels seen in the batch.
To compute those losses correctly, we design a custom sampler favoring consistent batches where the class prevalence is respected. 

\subsection{Injecting Prior Knowledge}
\begin{figure}[t]
\includegraphics[width=\textwidth]{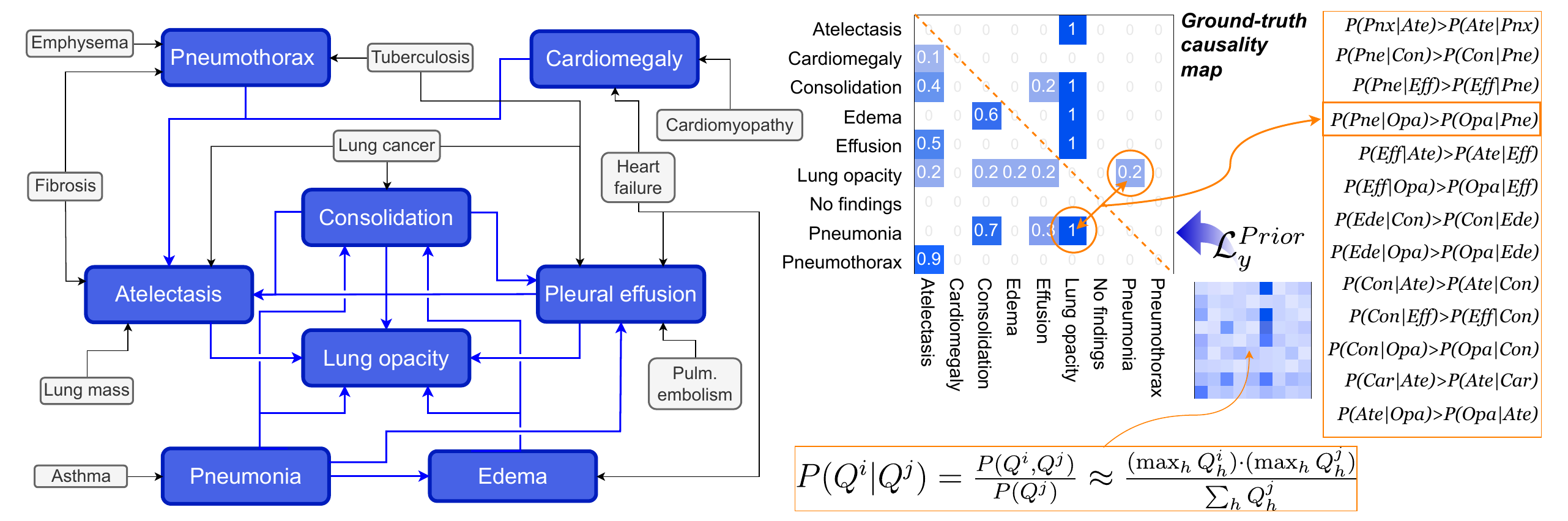}
\caption{Causal graphical model among the CXR findings of interest (blue) and the ground-truth \textit{causality map} defined from that graph. Gray boxes represent additional findings or risk factors (not investigated in this study) that might be associated with the desired ones.}
\label{fig:prior_knowledge}
\end{figure}

Motivated by the high interclass similarity and hierarchical structure of CXR findings \cite{rajaraman2020training,wang2017chestx}, we propose a new method to inject prior (medical) knowledge into the model to guide its learning (Fig.~\ref{fig:prior_knowledge}). Differently from solutions as \textit{conditional training} \cite{pham2021interpreting}, which rely on data, our proposal is desirable to capture semantic priors without relying on data.
We define a causal graph representing the relationship between the CXR findings and 
propose a novel formulation of the \textit{causality map} concept
 \cite{carloni2023causality,carloni2024exploiting} to model the co-occurrence of CXR findings in the images.
As we have seen, each $Q_y^{ca}$ representation has shape $n_c \times h$, where $n_c$ is the number of classes (e.g., nine CXR findings) and $h$ is the hidden dimension of the embeddings.
After normalizing $Q_y^{ca}$ by their global maximum batch-wise, they lie in the range $0$-$1$, and we interpret their values as probabilities of the CXR findings to be present in the image. Indeed, given two embeddings $Q^i$ and $Q^j$, to compute the effect of the former on the presence of the latter, we estimate the ratio between their joint and marginal probabilities as:
\begin{equation}
    P(Q^i|Q^j) = \frac{P(Q^i,Q^j)}{P(Q^j)} \approx \frac{(\max_{h} Q^i_{h})\cdot (\max_{h} Q^j_{h})}{\sum_{h} Q^j_{h}}, \forall i, j \in  1 \leq i,j \leq n_c   
\label{eq:cmap}
\end{equation}
\noindent thus obtaining the relationships between embeddings $Q^i$ and $Q^j$, since, in general, $P(Q^i|Q^j) \neq P(Q^j|Q^i)$. By computing these quantities for every pair $i,j$, we obtain the $n_c \times n_c$ map $C_y$. We interpret asymmetries across estimates opposite the main diagonal in $C_y$ as causality signals between their activation. Accordingly, the representation of a CXR finding causes the activation of another when $P(Q^i|Q^j) > P(Q^j|Q^i)$, that is $Q^i \rightarrow Q^j$.  
We design our \textbf{Task-Prior loss} as an MSE loss to push the causality map $C_y$ obtained from the learned representations to the ground-truth causality map $C_y^{GT}$ defined over CXR findings:
\begin{equation}
    \mathcal{L}_y^{Prior}= - \lambda_{12} (C_y-C_y^{GT})^2
    \label{eq:L_prior}
\end{equation}

Overall, the training objective of our CROCODILE framework is defined as the sum of the losses defined in Equations \ref{eq:L_y}, \ref{eq:L_d}, \ref{eq:L_RS}, \ref{eq:L_y_batch}, \ref{eq:L_d_batch} and \ref{eq:L_prior}:
\begin{equation}
    \mathcal L_{TOT} = \mathcal L_y + \mathcal L_d + \mathcal L_{RS} + \mathcal L_y^{batch} + \mathcal L_d^{batch} + \mathcal L_y^{prior}.
\end{equation}

\section{Experimental Setup}
We classify eight radiological findings (plus the \textit{No finding} class) from frontal CXR images of four popular data sets in both ID and OOD settings. After cleaning, the number of images for each set is: $112110$ for ChestX-ray14 \cite{wang2017chestx}, $183453$ for CheXpert \cite{irvin2019chexpert}, $95452$ for PadChest \cite{bustos2020padchest}, and $365737$ for MIMIC-CXR \cite{johnson2019mimic}.
For the first dataset, we create the \textit{Lung opacity} class as OR logic across the \textit{consolidation}, \textit{effusion}, \textit{edema}, \textit{pneumonia}, and \textit{atelectasis} classes.
We resize the images to $320 \times 320$ and adjust their contrast in $0$-$255$.
For ID experiments, we combine images of ChestX-ray14, CheXpert, and PadChest, split them into 80-20\% train and validation sets, and assess the multi-label classification performance via the area under the ROC curve (AUC) and the average precision (AP) scores for each category and their average.
We test the best-performing ID model on the external, never-before-seen MIMIC-CXR dataset to evaluate OOD generalization abilities.
In all the experiments, we adopted ResNet50 backbones, Adam optimizer, learning rate of 1e-6, batch size of $12$, and trained the model in early-stopping on an Nvidia A100 GPU with $64$ GB memory.
We compare to a regular ResNet50 architecture, a ResNet50 version of Nie \textit{et al}. \cite{nie2023chest} corresponding to discarding domain-branch and task-prior information from our method, our method without contrastive learning (CL) ($\mathcal L_{RS}$, $\mathcal L_y^{batch}$, $\mathcal L_d^{batch}$), and our method without the task prior (TP) ($\mathcal L_y^{prior}$).

\section{Results and Conclusion}
The results of our ID and OOD investigations (Table \ref{tab:results}) reveal our method is behind its ablated versions and \cite{nie2023chest} on i.i.d. data (ID) while is the best-performing model on the external never-before-seen data (OOD).
This important result points to a necessary trade-off between in-domain accuracy and out-of-domain robustness on real-world data, supporting recent work \cite{teney2024id}.
Notably, our method is the most effective in reducing the ID-to-OOD drop in performance.
By leveraging causal tools, disentanglement, contrastive learning, and prior knowledge, it learns a better mechanism from image to prediction, relies less on spurious correlations, and breaks the boundaries across domains. 
Our bias-mitigation proposal is general and can be applied to tackle domain shift bias in other computer-aided diagnosis applications, fostering a safer and more generalizable medical AI. 
\begin{table}[h]
    \centering
    \begin{tabular}{l|c|c|c|c|c}
        Finding & ResNet50 & Nie \textit{et al}. \cite{nie2023chest} & \textbf{Ours} \textit{w/o} CL & \textbf{Ours} \textit{w/o} TP & \textbf{Ours}\\
        \hline
        \multicolumn{6}{c}{\textbf{In-distribution} (ID) data}\\
        Atelectasis & 65.74/24.98 & 76.81/30.04 & \textbf{77.13}/30.26 & 77.07/30.37 & 77.04/\textbf{30.37}\\
        Cardiomegaly & 81.53/51.21 & 92.43/56.56 & \textbf{92.92}/\textbf{56.60} & 92.29/56.20 & 92.27/56.17\\
        Consolidation & 69.74/8.71 & 80.89/13.85 & 80.62/\textbf{14.10} & \textbf{81.13}/13.82 & 81.10/13.86\\
        Edema & 77.34/17.62 & 88.49/\textbf{23.01} & 88.21/22.53 & \textbf{88.73}/22.02 & 88.72/22.05\\
        Effusion & 77.69/51.26 & 88.68/56.31 & \textbf{89.08}/56.46 & 88.92/\textbf{56.65} & 88.93/56.65\\
        Lung opacity & 69.81/39.27 & 81.20/44.62 & \textbf{81.20}/\textbf{44.66} & 80.60/44.10 & 80.55/44.08\\
        No finding & 68.75/68.08 & \textbf{80.14}/73.46 & 79.68/\textbf{73.47} & 79.38/73.22 & 79.35/73.22\\
        Pneumonia & 67.76/20.74 & 78.05/\textbf{26.13} & \textbf{79.15}/25.73 & 77.65/24.86 & 77.63/24.85\\
        Pneumothorax & 78.86/32.78 & 89.87/\textbf{38.17} & \textbf{90.25}/37.69 & 88.79/37.02 & 89.86/37.03\\
        \textit{Mean} [$\uparrow$] & 73.02/34.96 & 84.06/\textbf{40.24} & \textbf{84.25}/40.17 & 83.95/39.81 & 83.94/39.81\\  
        \multicolumn{6}{c}{\textbf{Out-of-distribution} (OOD) data}\\
        Atelectasis & 62.79/31.56 & 74.02/36.69 & 74.11/36.63 & 74.15/\textbf{36.89} & \textbf{74.18}/36.83\\
        Cardiomegaly & 61.43/31.84 & 71.44/36.22 & 71.86/36.42 & \textbf{72.82}/37.16 & 72.80/\textbf{37.17}\\
        Consolidation & 66.41/7.20 & 77.01/11.97 & 77.38/\textbf{12.53} & 77.46/12.13 & \textbf{77.80}/12.07\\
        Edema & 74.04/36.12 & 84.52/40.48 & 83.95/40.46 & \textbf{85.43}/41.43 & 85.39/41.45\\
        Effusion & 75.10/59.66 & 86.16/64.60 & 86.04/64.87 & 86.01/64.85 & \textbf{86.49}/\textbf{64.99}\\
        Lung opacity & 56.92/28.52 & 67.86/33.49 & 67.43/33.10 & 68.30/33.83 & \textbf{68.31}/\textbf{33.85}\\
        No finding & 67.39/63.72 & 78.53/68.66 & 78.72/68.99 & \textbf{78.78}/69.02 & 78.74/\textbf{69.05}\\
        Pneumonia & 53.64/7.47 & 63.96/12.29 & 64.62/12.52 & 65.01/12.76 & \textbf{65.03}/\textbf{12.80}\\
        Pneumothorax & 64.72/12.39 & 74.89/16.76 & 75.41/17.65 & 75.48/17.70 & \textbf{76.11}/\textbf{17.72}\\
        \textit{Mean} [$\uparrow$] & 64.71/30.94 & 75.38/35.68 & 75.50/35.91 & 75.94/36.20 & \textbf{76.09}/\textbf{36.21}\\
        \hline
        ID-OOD drop& 11.38/11.50 & 10.33/11.33 & 10.38/10.60 & 9.54/9.07 & \textbf{9.35}/\textbf{9.04}
    \end{tabular}
    \caption{AUC/AP scores obtained on ID and OOD data. The drop is in percent.}
    \label{tab:results}
\end{table}
%
\newpage
%
%
%
\bibliographystyle{splncs04}
\bibliography{mybibliography}
\end{document}